\documentclass[aps,preprint]{revtex4}%
\usepackage{amsfonts}%
\usepackage{amsmath}%
\usepackage{amssymb}%
\usepackage{graphicx}%
\newtheorem{theorem}{Theorem}

\begin{document}
\preprint{}
\title{Factorization and escorting in the game-theoretical approach to non-extensive entropy measures}
\author{Flemming Tops\o e \thanks{Research supported by the Danish Natural Science
Research Council.}}
\affiliation{Department of Mathematics, University of Copenhagen, topsoe@math.ku.dk}
\keywords{Measure of complexity, factorization, escort probabilities}
\pacs{PACS numbers: 05.90.+m; 89.70.+c}

\begin{abstract}
  The game-theoretical approach to non-extensive entropy measures of
  statistical physics is based on an abstract \textit{measure of
    complexity} from which the \textit{entropy measure} is derived in
  a natural way. A wide class of possible complexity measures is
  considered and a property of factorization investigated. The
  property reflects a separation between the system being
  observed and the observer. Apparently, the property is also related
  to escorting. It is shown that only those complexity
  measures which are connected with \textit{Tsallis entropy} have the
  factorization property. 

\end{abstract}
\volumeyear{year}
\volumenumber{number}
\issuenumber{number}
\eid{identifier}
\date[Date text]{date}
\received[Received text]{date}

\revised[Revised text]{date}

\accepted[Accepted text]{date}

\published[Published text]{date}
\maketitle

\section{Entropy via complexity}

The present contribution should be seen in the context of the overall
goal: \textit{to give operational definitions of entropy and related
  quantities which cover the classical as well as non-extensive
  settings, thereby leading to an understanding of which of these
  quantities, in particular which entropy measures, are relevant for
  statistical physics.}

The findings here reported constitute a companion to the game
theoretical approach discussed in Tops\o e \cite{Topsoenext}. For the
convenience of the reader, we repeat some of the essentials.
The key objects of the approach are the \textit{alphabet} $A$, taken
to be discrete, the \textit{strategy set} $S_{I}$ of \textit{Player I}
(\textquotedblleft Nature\textquotedblright), also referred to as the
\textit{preparation}, a \textit{strategy set} $S_{II}$ for
\textit{Player II} (\textquotedblleft the physicist\textquotedblright)
and a real, or extended real valued \textit{complexity function}
$\Phi$ defined on the product set $S_{I}\times S_{II}$:
$(P,Q)\curvearrowright\Phi(P\Vert Q)$. The strategy set $S_{II}$ is
here taken to be the set of all probability distributions over $A$ and
$S_{I}$ is assumed to be a subset of $S_{II}$. A typical element in
$S_{I}$ is denoted $P$ and a typical element in $S_{II}$ is denoted
$Q$. Point probabilities are denoted by a subscript $i$ which then
ranges over $A$, e.g. $P=(p_{i})_{i\in A}$, $Q=(q_{i})_{i\in A}$.

The basic axiom we shall work with concerns the complexity function and states
that for each $P\in S_{I}$, the minimum over $Q\in S_{II}$ of $\Phi(P\Vert Q)$ is
finite and achieved for $Q=P$ and for no other probability distribution.

The \textit{entropy function} $S_{\Phi}$ is defined by
\begin{equation}
S_{\Phi}(P)=\inf_{Q\in S_{II}}\Phi(P\Vert Q)\,. \label{eq:1}%
\end{equation}
The essential requirement imposed on the complexity function is then
that the bi-implication
\begin{equation}
\Phi(P\Vert Q)=S_{\Phi}(P)\Leftrightarrow Q=P
\end{equation}
holds for any $P\in S_{I}$.
\textit{Divergence} $D(P\Vert Q)$ is the difference between complexity and
entropy:
\begin{equation}
D_{\Phi}(P\Vert Q)=\Phi(P\Vert Q)-S_{\Phi}(P)=\Phi(P\Vert Q)-\Phi(P\Vert P)\,.
\label{eq:2}%
\end{equation}

The game alluded to in the title is the two-person zero-sum game for which
Player I aims at high complexity and Player II at low complexity. As is
evident from \eqref{eq:1}, a \textit{maximum entropy distribution} is the same
as a \textit{maximin strategy}, i.e. as an \textit{optimal strategy} for
Player I. We are thus led directly to Jaynes \textit{maximum entropy
principle} ($MaxEnt$) and the game theoretical approach can to some extent be
said to explain that principle. A deeper understanding requires that also
optimal strategies for Player II are taken into consideration. For details,
see \cite{Topsoenext}.

We claim that the game theoretical view offers sound principles which
appear more natural and at a deeper level than $MaxEnt$ itself. It has
to be pointed out, however, that the approach does not give a clue to
which entropy measures appear naturally in physics. Some results which
address this problem from the game theoretical point of view are
contained in \cite{Topsoenext} and here we continue research in this
direction.  We focus on the triple $(\Phi,S,D)$ of complexity, entropy
and divergence and limit the study to triples of the form
\begin{align}
\Phi(P\Vert Q)  &  =\sum_{i\in A}\left(  q_{i}f(\frac{p_{i}}{q_{i}}%
)-f(p_{i})\right)  \,,\label{1}\\
S(P)  &  =-\sum_{i\in A}f(p_{i})\,,\label{2}\\
D(P\Vert Q)  &  =\sum_{i\in A}q_{i}f(\frac{p_{i}}{q_{i}})\,, \label{3}%
\end{align}
where $f$, the \textit{generator}, is a real-valued analytic and strictly
convex function on $[0,\infty\lbrack$ such that $f(0)=f(1)=0$ and such that
the normalization condition $f^{\prime}(1)=1$ is fulfilled. We may put
$f(\infty)=\infty$. The condition of analyticity, which does not appear in
\cite{Topsoenext}, guarantees that $f$ is smooth and that the behavior of $f$
in one interval, say in $[0,1]$, determines the behavior in the entire
interval $[0,\infty]$.

It is easy to check, by appeal to Jensen's inequality, that with $S$
and $D$ given by \eqref{2} and \eqref{3}, $S=S_{\Phi}$ and
$D=D_{\Phi}$.

Often, it is advantageous to work with the \textit{dual generator}
$\widetilde f$.  This is the function given by
\begin{equation}
\widetilde{f}(x)=xf\left(  \frac{1}{x}\right)  \,,\,\,\,0\leq x\leq\infty\,.
\end{equation}

The formulas \eqref{1}-\eqref{3} expressed in terms of the dual generator give
the expressions
\begin{align}
\Phi(P\Vert Q)  &  =\sum_{i\in A}p_{i}\left(  \widetilde{f}\left(  \frac{q_{i}%
}{p_{i}}\right)  -\widetilde{f}\left(  \frac{1}{p_{i}}\right)  \right)
\,,\label{4}\\
S(P)  &  =-\sum_{i\in A}p_{i}\widetilde{f}\left(  \frac{1}{p_{i}}\right)
\,,\label{5}\\
D(P\Vert Q)  &  =\sum_{i\in A}p_{i}\widetilde{f}\left(  \frac{q_{i}}{p_{i}%
}\right)  \,. \label{6}%
\end{align}

We realize that a very wide class of potentially interesting entropy measures
are of the form here considered. In the classical case, leading to the
\textit{Boltzmann-Gibbs-Shannon entropy}, we have $f(x)=x\ln x$ and
$\widetilde{f}(x)=\ln\frac{1}{x}$. Further cases work with \textit{deformed
logarithms}. A two-parameter version of these, taken from \cite{Topsoenext},
are given by
\begin{equation}
\ln_{\alpha,\beta}x=%
\begin{cases}
\frac{x^{\beta}-x^{\alpha}}{\beta-\alpha}\mbox{ if
}\beta\not =\alpha\,,\\
x^{\alpha}\ln x\mbox{ if }\beta=\alpha\,.
\end{cases}
\label{7}%
\end{equation}
With appropriate choice of parameters ($-1<\alpha\leq0$, $0\leq\beta$ or the
symmetric choice) this leads to the family of generators given by
\begin{equation}
f_{\alpha,\beta}(x)=x\ln_{\alpha,\beta}(x)\mbox{ with dual }\widetilde
{f}_{\alpha,\beta}(x)=\ln_{\alpha,\beta}\frac{1}{x}\,. \label{9}%
\end{equation}
With $(\alpha,\beta)=(1-q,0)$ for a positive value of the parameter $q$, we
are led to \textit{Tsallis $q$-entropy}, cf. Tsallis
\cite{Tsallis88}. Listing also the associated complexity and divergence
functions we find that they are given by the expressions:
\begin{align}
\Phi_{q}(P\Vert Q)  &  =\frac{1}{1-q}\sum_{i\in A}p_{i}^{q}\left(  1-q_{i}%
^{1-q}\,\right)  ,\label{10}\\
S_{q}(P)  &  =\frac{1}{1-q}\sum_{i\in A}p_{i}\left(  p_{i}^{q-1}-1\right)
=\frac{1}{1-q}\left(  \sum_{i\in A}p_{i}^{q}-1\right)  \,,\label{11}\\
D_{q}(P\Vert Q)  &  =\frac{1}{1-q}\sum_{i\in A}p_{i}\left(  1-\left(
\frac{p_{i}}{q_{i}}\right)  ^{q-1}\right)  =\frac{1}{1-q}\left(  1-\sum_{i\in
}p_{i}\left(  \frac{p_{i}}{q_{i}}\right)  ^{q-1}\right)  \,. \label{12}%
\end{align}

A significant feature is the \textit{factorization property} of \eqref{10}.
Note that this structural property is not immediately visible just looking at
the entropy function \eqref{11}.

\section{Factorization}

Let us have a closer look at the factorization property and
investigate for which generators $f$ the associated complexity
function factorizes.  In more detail, with $\Phi$ of the form
\eqref{1}, what we search for are functions $\xi$ and $\zeta$ such
that
\begin{equation}
\label{13}\Phi(P\Vert Q)=\sum_{i\in A}\xi(p_{i})\zeta(q_{i})
\end{equation}
holds generally. We call $\xi$ the \textit{escort function}. It acts
as a kind of \textit{mean-value generator} and invites for
\textit{escorting}, which is the consideration, along with any
distribution $P=(p_i)_{i\in A}$ of the associated \textit{escort
  distribution} with point probabilities $(\xi(p_i)/Z)_{i\in A}$ where
$Z$ is the appropriate constant of normalization. The consideration of
escort probabilities occurs in connection with $MaxEnt$ but presents
in itself problems regarding the proper physical interpretation. The
literature can be traced from the recent contribution by Ferri,
Mart\'{i}nez and Plastino, cf.  \cite{Ferrietal05} and, for critical
comments, from Shalizi \cite{Shalizi05}. The function $\zeta$ is the
\textit{surprise factor}.  This may also be thought of as
\textit{self-information}.  We require that $\xi(1)=1$ and
$\zeta(1)=0$ and that both functions are analytic in $[0,\infty[$.
Thus, we demand that escorting does not change a deterministic
distribution and that the surprise value associated with the
occurrence of a deterministic event is $0$. The escort function and
the surprise function need in fact only be given for arguments in
$[0,1]$.  However, the requirement of analyticity is a requirement
that these functions have natural extensions to functions defined for
any positive argument.  Therefore, we may as well assume that the
functions are defined from the outset on the whole interval
$[0,\infty[$.

Now assume that $\Phi$ given by \eqref{1} factorizes as described
above.  Then, from the formula
\begin{align*}
D(P\Vert Q)  &  =\Phi(P\Vert Q)-\Phi(P\Vert P)=\sum_{i\in A}\xi(p_{i})\left(
\zeta(q_{i})-\zeta(p_{i})\right) \\
&  =\sum_{i\in A}p_{i}\widetilde{f}\left(  \frac{q_{i}}{p_{i}}\right)\,,
\end{align*}
we realize that
\begin{equation}
\xi(x)\left(  \left(  \zeta(y)-\zeta(x)\right)  \right)  =x\widetilde
{f}\left(  \frac{y}{x}\right)  \,. \label{14}%
\end{equation}

In fact, by simple considerations, also using a continuity argument, this is
first established for $x$ and $y$ in the unit interval. The general validity
then follows by analyticity considerations.

Putting first $x=1$, then $y=1$ in \eqref{14} we realize that $\xi$ and
$\zeta$ can be identified in terms of $\widetilde{f}$. In fact,
\begin{equation}
\zeta=\widetilde{f}\mbox{ and }\xi(x)=\frac{x\widetilde{f}(\frac{1}{x}%
)}{-\widetilde{f}(x)}\,. \label{15}%
\end{equation}
Inserted into \eqref{14} we see that
\[
\frac{x\widetilde{f}(\frac{1}{x})}{-\widetilde{f}(x)}\left(  \widetilde
{f}(y)-\widetilde{f}(x)\right)  =x\widetilde{f}\left(  \frac{y}{x}\right)
\,,
\]
hence
\[
\widetilde{f}(\frac{1}{x})\left(  \widetilde{f}(x)-\widetilde{f}(y)\right)
=\widetilde{f}(x)\widetilde{f}\left(  \frac{y}{x}\right)
\]
or
\[
\widetilde{f}(\frac{1}{x})\left(  \widetilde{f}(x)-\widetilde{f}(xy)\right)
=\widetilde{f}(x)\widetilde{f}(y)\,.
\]
Then
\begin{align*}
\frac{\widetilde{f}(x)+\widetilde{f}(y)-\widetilde{f}(xy)}{\widetilde
{f}(x)\widetilde{f}(y)}  & =\frac{1}{\widetilde{f}(x)}+\frac{\widetilde
{f}(x)-\widetilde{f}(xy)}{\widetilde{f}(x)\widetilde{f}(y)}\\
&  =\frac{1}{\widetilde{f}(\frac{1}{x})}+\frac{1}{\widetilde{f}(x)}\\
&  =\alpha\,,
\end{align*}
a constant (because of the dependence on $x$ only in the last expression and
the symmetry in $x$ and $y$ of the expression we started out with). In case
$\alpha=0$ we are soon led to conclude that then $\widetilde{f}(x)=\ln
\frac{1}{x}$ and we are faced with the classical Boltzmann-Gibbs-Shannon
case. When $\alpha\not =0$, we use the functional equation
\[
\left(  1-\alpha\widetilde{f}(xy)\right)  =\left(  1-\alpha\widetilde
{f}(x)\right)  \left(  1-\alpha\widetilde{f}(y)\right)
\]
and conclude, as $\widetilde{f}^{\prime}(1)=-1$, that $1-\alpha\widetilde
{f}(x)=x^{\alpha}$, hence
\[
\widetilde{f}(x)=\frac{1-x^{\alpha}}{\alpha}=\ln_{-\alpha,0}\frac{1}{x}\,.
\]

As $f$, hence also $\widetilde{f}$, is required to be convex, we must
conclude that $\alpha<1$. We realize that we are faced with the
Tsallis family (with parameter $q=1-\alpha$).

To sum up, we have proved:

\begin{theorem}
  A complexity function of the form \eqref{1} factorizes if and only
  if it is related to one of the Tsallis entropies,
  cf. \eqref{10}-\eqref{11}.
\end{theorem}

\section{Conclusions}

Factorization of the complexity function appears important as it
separates the two sides of the game, that of nature and that of the
physicist. The property can be used to characterize the family of
entropy measures suggested by Tsallis. 

Though our results are, hopefully, indicative of central issues for
fundamentals of non-extensive statistical physics, operational
interpretations are not yet in place.  The game theoretical approach
may well be helpful in this connection, see \cite{Topsoenext} for some
further results, especially on equilibrium, and for open problems to
look into.  


\begin{thebibliography}{1}

\bibitem{Ferrietal05}
G~L Ferri, S~Mart\'{i}nez, and A~Plastino.
\newblock Equivalence of the four versions of Tsallis's statistics.
\newblock {\em Journal of Statistical Mechanics: Theory and Experiment},
  2005(04):P04009, 2005.

\bibitem{Shalizi05}
C.~R. Shalizi.
\newblock Tsallis {S}tatistics, {S}tatistical {M}echanics for {N}on-extensive
  {S}ystems and {L}ong-{R}ange {I}nteractions.
\newblock Available at http://cscs.umich.edu/~crshalizi/notebooks/tsallis.html.

\bibitem{Topsoenext}
F.~Tops{\o }e.
\newblock Entropy and {E}quilibrium via {G}ames of {C}omplexity.
\newblock {\em Physica A}, 340/1-3:11--31, 2004.

\bibitem{Tsallis88}
C.~Tsallis.
\newblock Possible generalization of {B}oltzmann-{G}ibbs statistics.
\newblock {\em J. Stat. Physics}, 52:479, 1988.
\newblock {S}ee http://tsallis.cat.cbpf.br/biblio.htm for an extensive
and updated bibliography.

\end{thebibliography}

\end{document}